\documentstyle[preprint,aps]{revtex}

\renewcommand{\imath}{i}

\def\bml{\begin{mathletters}}
\def\eml{\end{mathletters}}
\begin{document}
\title{Quantum Coherence in an Exactly Solvable One-dimensional Model with
Defects}
\author{ {\sc P.~Schmitteckert$(^*)$,
P.~Schwab} and {\sc U. Eckern}\\
 {\em Institut f\"ur Physik, Universit\"at Augsburg}\\
 {\em D-86135 Augsburg, Germany}
}
\date{18.5.1995}
\maketitle
\begin{abstract}
{\em Abstract}. --
Using the Quantum Inverse Scattering Method we construct an integrable
Heisen\-berg-XXZ-model, or equivalently a model for spinless fermions with
nearest-neighbour interaction,
with defects.
Each defect involves three sites with a fine tuning between nearest-neighbour
and next-nearest-neighbour
terms. We investigate the finite size corrections to the ground state energy
and its dependence on an external flux
as a function of a parameter $\nu$, characterizing the strength of the defects.
For intermediate values of $\nu$, both quantities become very small, although
the ground state wavefunction
remains extended.\smallskip
\par\noindent
$(^*)$ Electronic mail: Peter.Schmitteckert@physik.uni-augsburg.de
\end{abstract}
\ \vspace{3ex} \\
\noindent
\begin{tabular}{ll@{\ --- \ }l}
 PACS: & 71.27.+a & Strongly correlated electron systems.\\
       & 71.30.+h & Metal-insulator transitions.\\
       & 75.10.Jm & Quantized spin models.
\end{tabular}
\ \vspace{9ex}\\
\noindent
{\em Introduction}. --
Three recent experiments have demonstrated that persistent currents, periodic
in the
magnetic flux, exist in mesoscopic metal \cite{Levy} and
semiconductor \cite{Benoit93} rings at very low temperatures.
Surprisingly, though the current is found to be small, of the order of $\sim
ev_F/L$
for single rings ($v_F$ is the Fermi velocity, and $L$ the circumference), it
is still two orders
of magnitude larger than expected theoretically, at least for the metal rings
studied in \cite{Levy}.
In the latter, the electron motion is diffusive, i.e. the elastic mean free
path is much smaller than the circumference. While it is well established that
the Coulomb interaction gives an important contribution to the current for a
measurement on
an ensemble of rings \cite{Ambegaokar91},
the interaction effect in single rings, is far from being understood
theoretically.\par
In this article, we  consider a one-dimensional,
interacting model in the presence of a magnetic flux, or equivalently, with
twisted boundary conditions.
We introduce very special
``defects'' into the model describing spinless fermions with nearest-neighbour
interaction.
Despite this inhomogeneity, the model remains integrable and we present exact
results for the finite size corrections
to the ground state energy, and its dependence on the magnetic flux, as a
function of a parameter $\nu$
characterizing the strength of the defects.
Clearly, our investigation does not provide an answer to the
questions raised by the experiments (there, the number of transverse channels
is much larger than one).
Instead, our work is closely related to, and an extension of, various recent
theoretical studies
[4--9]
of quantum coherence in strongly interacting electron systems.\par
\noindent
{\em Construction of the model}. --
Using the Quantum Inverse Scattering Method (QISM), we construct our model from
the
${\cal R}$ and ${\cal L}$ matrices of the  Heisenberg-XXZ-model on an
inhomogeneous lattice as, for example, described
in \cite{Korepin93}. The central equation of the QISM  is the Yang-Baxter
equation, which guarantees that a
scattering process factorizes in two-particle scattering processes and does not
depend on the order of these.
In order to construct a model with defects, we allow that the local ${\cal
L}_n$ matrix depends,
in addition to the spectral parameter $\lambda$,  on a parameter $\nu_n$,
${\cal L}_n(\lambda) = {\cal L}(\lambda+\nu_n)$.
The transfer matrix is given by $T(\lambda)= \mbox{Tr} \prod_{n=1}^{M} {\cal
L}(\lambda + \nu_n)$,
where $M$ denotes the number of lattice sites.
To include twisted boundary conditions,  we multiply the ${\cal L}_M$ matrix of
the  Heisenberg-XXZ-ring
with  $\exp{\left( \imath \phi\,\hat{\sigma}^z /2\right)}$.\par
The Hamiltonian is then given as the logarithmic derivative of the transfer
matrix with respect to $\lambda$,
at a specific value \cite{Korepin93}.
In particular, in the special case in which all $\nu_n=0$, we obtain the usual
XXZ-model, which can
be transformed to a spinless fermion model by a Jordan-Wigner transformation.
For a general set of parameters, $\{\nu_n\}$,  it is difficult to determine the
Hamiltonian explicitly,
with one exception, namely where there are no defects on neighbouring sites,
i.e. $\nu_n\nu_{n+1}=0$ for all $n$.
This is the situation we study in the following.
As an illustration, consider a vanishing nearest-neighbour interaction, and a
single defect at the site $n_1$
characterized by the parameter $\nu$. The resulting Hamiltonian is given by
\begin{eqnarray}
  {\cal H} &=& {\cal H}^0 \;+\;{\cal H}^I_{n_1}(\nu) \;=\;
       - \sum_{n=1}^M \Big( c^{+}_n c^{}_{n+1} \;+\;  c^{+}_{n+1}
c^{}_{n}\Big)\;+\; {\cal H}^{I}_{n_1}(\nu)\\
  {\cal H}^I_{n_1}(\nu) &=& \big( 1 - \frac{1}{\cosh \nu}\Big)\,\Big(
c^{+}_{n_1-1} c^{}_{n_1} \;+\; c^{+}_{n_1} c^{}_{n_1+1} \Big)
  \;-\; e^{\imath\pi/2}\,\tanh(\nu)\,  c^{+}_{n_1-1} c^{}_{n_1+1}\;+\;
\mbox{h.c.} \label{def:HI}
\end{eqnarray}
where the $\{c^{+}_n\}$ and $\{c^{}_n\}$ are the standard fermion creation and
annihilation operators.
The generalization to $r$ defects is straightforward (assuming
$\nu_n\nu_{n+1}=0$),
\begin{equation}
{\cal H}={\cal H}^0+\sum_{\ell=1}^{r} {\cal H}^I_{n_\ell}(\nu_{n_\ell})
\end{equation}
where $n_\ell$ denotes the location of a defect with strength $\nu_{n_\ell}$.
An illustration is given in Fig. \ref{fig:Defect}.
The expression for the Hamiltonian in the presence of a finite
nearest-neighbour interaction
is more lengthy but similar in structure, i.e. a defect located at $n_\ell$
affects the lattice sites
$n_\ell-1$, $n_\ell$, and $n_\ell+1$ only \cite{Schmitteckert}.
\par\noindent
{\em Single defect, no interaction}. --
As is apparent from Eq. (\ref{def:HI}), for $\nu=0$, the Hamiltonian reduces to
${\cal H}^0$, i.e. the standard single-band
tight-binding model (the hopping matrix element is chosen to be unity). In the
opposite limit, $\nu=\infty$,
the lattice site $n_1$ is cut out of the ring. As a result, the model
represents free fermions on a
ring of $M-1$ sites,
however, with an additional phase factor $e^{\imath\delta_1}$,
$\delta_1=\pi/2$, for the hopping between $n_1-1$ and
$n_1+1$, plus one uncoupled site.
We emphasize that the parameters, $\cosh^{-1}(\nu)$ and $\tanh{(\nu)}$, as well
as the phase factor
$\delta_1=\pi/2$, are fine-tuned in the following sense:
a generic impurity breaks translational invariance and lifts the degeneracies
of the single-particle spectrum, which are found at $\phi=0,\pm\pi$. While our
defects also break translational invariance, this symmetry is replaced by
another,
of not as clear physical origin. As a result we find that even when changing
$\nu$, no degeneracies
are lifted ---
they only occur at different, $\nu$-dependent values of $\phi$.
The corresponding $\nu$-dependent symmetry operators can be constructed
\cite{Korepin93}.\par
The  localization of electronic states is another, well established phenomenon,
in one-dimensional disordered systems.
In Fig. \ref{plot:WF}, we plot
the squared modulus of the wavefunction for the single-particle level lowest in
energy.
Clearly, for the integrable case,
the wavefunction is extended though reduced at the defect. Allowing, however,
the phase $\delta_1$ to be different from $\pi/2$, which corresponds to the
non-integrable case, we find a drastically
different behaviour with a clear localization of the wave-function near the
defect.
\par
{\em Several defects, finite interaction}. --
The results for a finite nearest-neighbour interaction, and in the presence of
several defects, are obtained from the
Bethe equations, which we derive from the algebraic Bethe ansatz, with the
result
\begin{equation} \label{eq:BE}
  \left[\frac{ \cosh(\lambda_j-\imath\eta)}{\cosh(\lambda_j+\imath\eta)}
   \right]^{M-r}
    \prod^{r}_{\ell=1}
    \frac{\cosh(\lambda_j+\nu_{n_\ell}-\imath\eta)
     }{\cosh(\lambda_j+\nu_{n_\ell}+\imath\eta)} \;=\; e^{\imath\phi}\,
     \prod^{N}_{ k=1 \atop k\ne j}
     \frac{\sinh(\lambda_j-\lambda_k - 2\imath\eta)
     }{\sinh(\lambda_j-\lambda_k + 2\imath\eta)}.
\end{equation}
Here,  $N$ is the number of fermions and $M$ and $r$ denote the number of
lattice sites and defects, respectively.
The nearest-neighbour interaction, in units of the hopping matrix element,
is parametrized as $\Delta=\cos{(2\eta)}$ (we consider $-1 < \Delta < 1$ only,
$\Delta>0$
corresponds to an attractive interaction).
For the spin model and for the fermion model with an odd number of fermions,
$\phi=0$ ($\pi$) represents
periodic (antiperiodic) boundary conditions, and vice versa for an even number
of fermions.
The phase $\phi$ is directly related to the magnetic flux $\Phi$, for odd $N$
according to the relation $\phi = 2\pi \Phi/\Phi_0$,
 $\Phi_0 = h/e$. (For even $N$, $\phi = 2\pi \Phi/\Phi_0 -\pi$.)
\par
The equations  (\ref{eq:BE}) have a remarkable consequence: given a set of
defect parameters $\{\nu_n\}$
the roots $\{ \lambda_j\}$ of the Bethe equations do not depend on the
distribution of the defects
over the sites.
This implies that the defects can be moved or permutated
without any change in the energy spectrum,
which is quite different from the observation that in mesoscopic physics,
thermodynamic and transport
properties can vary considerably by moving only one impurity by a few lattice
constants.\par
In the following, we simplify our defect model further, by assuming that
$|\nu_{n_\ell}|=\nu$, and choosing an equal number of
defects of strength $+\nu$ and $-\nu$. With this choice, the ground state
energy is an even function of the phase $\phi$.
We define $x= \lim_{r,M\rightarrow\infty} (r/M)$ to be the density of defects.
Assuming in addition
$N=M/2$, i.e. a half filled band,
we calculate in a first step the ground state energy per lattice site in the
thermodynamic limit,
$e_\infty = \lim_{M\rightarrow\infty} e_M$, $e_M \equiv E_M/M$.
Using the standard method \cite{Korepin93,YangYang66}, we find
\begin{eqnarray}
  e_\infty &=&  -\sin^{2}(2\eta) \int^{\infty}_{-\infty}  \frac{
\rho_{\infty}(\lambda) }{\cosh(\lambda+\imath\eta)\cosh(\lambda-\imath\eta)}
\mbox{d}\!\lambda \;-\;\frac{\cos(2\eta)}{2} \label{eq:ETDL}\\
 \rho^0_\infty(\lambda)&=& \Big\{\, 2(\pi-2\eta)\,\cosh [
\pi\lambda/(\pi-2\eta) ] \,\Big\}^{-1}\\
   \rho_\infty(\lambda) &=& (1-x) \rho^0_\infty(\lambda) \;+\; x\,\Big[
\rho^0_\infty(\lambda+\nu)\;+\; \rho^0_\infty(\lambda-\nu)\Big]/2.
\label{eq:rhoinfty} \label{eq:rhoinftynull}
\end{eqnarray}
The last equation relates the density of roots for the infinite system,
$\rho_\infty$, to the corresponding density of
roots of the homogeneous system, $\rho^0_\infty$ .
For $\nu=0$, clearly, $\rho_\infty=\rho^0_\infty$, and we recover the results
of \cite{YangYang66}.
On the other hand, for $\nu\rightarrow\infty$, it follows from
(\ref{eq:rhoinftynull}) that
$\rho_\infty = (1-x) \rho^0_\infty$; as a direct consequence, the integral in
Eq. (\ref{eq:ETDL}) is simply multiplied by the factor
$(1-x)$ compared to the homogeneous case, and $e_\infty$ is easily calculated.
Apparently, for all values of $\nu$, the ground state energy $e_\infty$ is a
linear function of the defect concentration
(and it is independent of $\phi$, i.e. the boundary conditions).\par
The leading order finite size corrections can be determined with the help of
the Wiener-Hopf technique, as described in
\cite{Hamer87}, with the result
\begin{eqnarray} \label{eq:FSC}
  e_M(\phi) - e_\infty&=&-\frac{(\pi^2/6) \sin{(2\eta)}/(\pi-2\eta)}{1-x +
x\cosh{[\pi \nu/(\pi-2\eta)]}}
  \left( 1 - \frac{ 3 \phi^2}{4\pi\eta} \right) \frac{1}{M^2},
\end{eqnarray}
which for $x=0$ is in agreement with \cite{Hamer87}.
In contrast to $e_\infty$, the finite size corrections depend in a nonlinear
way on the defect density,
and they are exponentially suppressed,
$\sim \exp{[ -\pi\nu/(\pi-2\eta)]}$, in the limit of large $\nu$.
 The finite size corrections, $(e_M-e_\infty)M^2$, obtained by solving
numerically the Bethe equations (\ref{eq:BE}), are plotted in Fig.
\ref{plot:FSC} as a function of the defect parameter $\nu$,
for half filling ($N=M/2$), and $\Delta=0.5$, $x=0.2$. The system size is
varied from $10^2$ to $10^4$.
The numerical result is in perfect agreement with the analytical expression
(\ref{eq:FSC}), but we also observe that the
limits $M\rightarrow\infty$ and $\nu\rightarrow\infty$ do not commute.
Taking   $\nu\rightarrow\infty$ first, we obtain a system with $r$ sites cut
out; however,
the occupancy of these sites still appears in the Hamiltonian, but as a good
quantum number.
The finite size corrections thus correspond to a system of $M-r$ sites, i.e.
they are enhanced in magnitude by the factor
$(1-x)^{-1} = 5/4$ as compared to a system with $x=0$. For large values of
$\nu$, the asymptotic result (\ref{eq:FSC})
(applicable for $\nu$ finite, $M\rightarrow\infty$) is only reached for
extremely large systems.\par
Finally, we consider the phase sensitivity of the ground state energy, similiar
to Refs.  [4-6],
where the homogeneous system was studied.
Here we choose $M=10^3$ and $x=0.2$, and we plot $[e_M(\pi)-e_M(0)]M^2$ as a
function of $\nu$ in Fig. \ref{plot:PS} for different fillings $N/M$.
Close to half filling ($N=440, 490$), the dependence on $\nu$ is very similar
to the result described
above for the finite size
corrections, i.e. the phase sensitivity is very small between $\nu\approx4$ and
$\nu\approx8$.
In this context, note that Eq. (\ref{eq:FSC}) implies that
$[e_M(\pi)-e_M(0)]/[e_M(0)-e_\infty] = -3\pi/(4\eta)$; but note also
that different $\Delta$'s have been chosen in Figs. \ref{plot:FSC} and
\ref{plot:PS}.\par
The asymptotic (large $\nu$) results shown in Fig. \ref{plot:PS} can be
explained as follows:
in this limit (and for the given parameters), 200 sites are cut
out from the system of 1000 sites. But, as discussed above, the occupancy of
these sites
still enters into the Hamiltonian.
The defect phases, $\delta_\ell$, are found to be $\pm 2\eta$, and as a result,
we find
additional cusps in the energy-phase relation, for example, for $\eta$ slightly
less than $\pi/4$, at $\phi=\pm 4 \eta$.
Consequently, the number of discontinuities in the persistent current,
$I=-\partial E_M/\partial\Phi$, increases, implying a reduction of the phase
average of the current over half a period,
i.e. $e_M(\pi)-e_M(0)$, compared to its $\nu=0$ value.
\par
For low filling, $N\ll M$, on the other hand, the interaction is unimportant
and we may apply the free electron result,
$I=-(ev_F/L)\phi/\pi$, which implies that $E_M(\pi)-E_M(0) \approx \pi^2N/M^2$
for small $N$.
(In our units, $L=M$ and $\hbar v_F=2 \sin(\pi N/M)$.)
Comparing the limits $\nu=0$ and $\nu\rightarrow\infty$, the phase sensitivity
is thus approximately enhanced in the latter case by the factor $(1-x)^{-2}$,
which is
apparent in Fig. \ref{plot:PS}.
\par\noindent
{\em Conclusions}. --
We have presented exact results for the finite size corrections to the ground
state energy and its flux sensitivity
for an one-dimensional, interacting model in the presence of defects.
Through its construction, the model remains integrable.
The defects are ``magnetic'' in the sense that a defect triangle, compare Fig.
\ref{fig:Defect}, encloses a finite magnetic
flux.
This means that time reversal symmetry is broken and, for general $\{ \nu_n\}$,
the energy is not an even function of $\phi$.
Surprisingly, the energy spectrum is independent of the spatial distribution of
defects,
and we find neither level repulsion nor localized states, which are considered
to be generic properties of a ``real'' impurity.
We believe that the absence of these effects is strongly related to the
integrability of the model.
Nevertheless, the finite size corrections and the phase sensitivity, i.e. the
persistent current,
can become very small.
\par
\begin{center}
 ***
\end{center}
\par
We thank {\sc K.-H. H\"ock} for interesting discussions and useful comments on
the manuscript.
P.~Schmitteckert acknowledges financial support through a graduate student
fellowship
by the state of Bavaria.
\par
\renewcommand{\labelenumi}{(\alph{enumi})}

\par\vspace{2ex}\par
\newpage
{\center{\large\bf List of Figures}}
\begin{description}
 \item[Figure \ref{fig:Defect}:] \refstepcounter{figure} \label{fig:Defect}
  Graphical representation of a defect at the site $n_\ell$, compare Eq.
(\ref{def:HI}). Note that for $\nu \rightarrow 0$
 ($\nu  \rightarrow \infty$) the dashed (dotted) lines representing the
corresponding hopping contributions are effectively cut.
 \item[Figure \ref{plot:WF}:] \refstepcounter{figure}  \label{plot:WF}
 Squared modulus of the wavefunction of the lowest energy eigenstate for the
non-interacting limit
 in the presence of a single defect ($\nu=1$) at the site $n_1=50$ ($M=100$).
 For the integrable case ($\delta_1=\pi/2$), the eigenstate is extended, though
reduced at the defect,
 while a detuning of the defect phase
  ($\delta_1\ne \pi/2$) leads to a localization of the wavefunction.
 \item[Figure \ref{plot:FSC}:] \refstepcounter{figure}   \label{plot:FSC}
 Finite size corrections to the ground state energy, ($e_M - e_\infty)\,M^2$,
as a function of
 the defect parameter $\nu$ for half filling, $N=M/2$.
 The system size is varied from $M=10^2$ to $M=10^4$; the results for large
systems fit perfectly
 with the analytical expression, Eq. (\ref{eq:FSC}) for $\phi=0$.
 The nearest-neighbour interaction is chosen to be $\Delta=0.5$, and the defect
density is $x=0.2$.
 \item[Figure \ref{plot:PS}:] \refstepcounter{figure}  \label{plot:PS}
 Phase sensitivity of the ground state energy, $[ e_M(\phi=\pi) - e_M(0) ]
M^2$, as a function of $\nu$ for
 different fillings, i.e. numbers of fermions, $N$ ($M=10^3$, $\Delta=0.2$).
\end{description}
\end{document}